\documentclass[twocolumn,showpacs,prl,aps,amssymb,superscriptaddress]{revtex4}
\usepackage{graphicx}
\usepackage[ansinew]{inputenc}
\usepackage{array}
\usepackage{color}
\usepackage{amsmath}
\usepackage{amsxtra}
\usepackage{amstext}
\usepackage{amssymb}
\usepackage{latexsym}
\usepackage{dsfont}
\usepackage{verbatim}
\usepackage{comment}

\renewcommand{\vec}[1]{{\bf{#1}}}

\begin{document}
\title
{Superradiance induced topological vortex phase in a Bose-Einstein condensate}

\author{M.E. Ta\c{s}g\i n}
\affiliation{College of Optical Sciences and Department of Physics, The University of Arizona, Tucson, Arizona 85721, USA}

\author{\"{O}.E. M\"{u}stecapl\i o\~{g}lu}
\affiliation{Department of Physics, Ko\c{c} University, 34450 Sar{\i}yer, Istanbul, Turkey}
  
\author{L. You}
\affiliation{Department of Physics, Tsinghua University, Beijing
100084, P. R. China}

\date{\today}

\begin{abstract}

We investigate theoretically a topological vortex phase transition induced by a superradiant  phase transition in an atomic  Bose-Einstein condensate driven by a Laguerre-Gaussian optical mode. We show that superradiant radiation can either carry zero angular momentum, or be in a rotating Laguerre-Gaussian mode with angular momentum. The conditions leading to these two regimes are determined in terms of the width for the pump laser and the condensate size for the limiting cases where the recoil energy is both much smaller and larger than the atomic interaction energy.

\end{abstract}
\maketitle

In addition to the spin angular momentum (SAM) associated with its polarization \cite{Beth}, electromagnetic radiation can also carry orbital angular momentum (OAM) associated with its spatial mode structure~\cite{cohen} as for example in Laguerre-Gaussian (LG) beams \cite{AllenPRA1992}. It can exhibit quantum entanglement between OAM states~\cite{ZeilingerNature2001}, and it was proposed~\cite{MolinaPRL2002} and demonstrated~\cite{MolinaPRL2004-MolinaNaturePhys2007} that superpositions of photonic OAM states can be utilized for higher-dimensional quantum communication and in dense data storage applications~\cite{KazslikowskiPRL2000,CollinsPRL2002}.

The efficient processing and storage of quantum information in terms of OAM requires reliable mechanisms for exchanging angular momentum between photons and atoms. For a superfluid state such as an atomic Bose-Einstein condensate (BEC), OAM states translate into topological excitations~\cite{TabosaPRL1999,AndersonPRL2006,WrightPRA2008,MorettiPRA2009}, e.g. as vortices or circulating modes on a surface \cite{PethickSmith}. The transfer of angular momentum involving LG beam induced vortices in atomic condensates have been studied both experimentally~\cite{TabosaPRL1999,AndersonPRL2006,WrightPRA2008,MorettiPRA2009} and theoretically~\cite{MarzlinPRL1997,MeystrePRA2007,DuttonPRL2004-ThanvanthriPRA2008}. The present paper reports conditions under which that transition can occur in a BEC simultaneously with the onset of a topological vortex. 

Under appropriate conditions, an ensemble of atoms optically driven above a threshold intensity can emit radiation in the form of a superradiant pulse (SR) \cite{dicke,skribanowitz}, a process analogous to a first-order phase transition (PT)~\cite{WangPRA1973-heppAnnPhys,EmaryPRE2003}. Early work on BEC superradiance~\cite{ketterle,ketterleRaman,bonifacioexp,JuntaoLiPLA2008,Deng} analyzed it in terms of matter-wave gratings in the translational \cite{ketterle,MeystrePRL1999-MustecapPRA2000,zobaySpatial,bonifacioexp,JuntaoLiPLA2008,Deng} and in the polarization~\cite{ketterleRaman,meystreRamanSR} degrees of freedom of the atoms. More recently several authors also discussed and demonstrated the onset of a structural phase transition from a homogenous to supersolid phase in a BEC trapped in an optical cavity ~\cite{EsslingerNature2010,NagyEPJD2008-DomokosPRL2002}. In contrast,  the present situation involves gratings in both translational and rotational degrees of freedom. Above the SR threshold, collective scattering of an incident LG pump laser results in the sudden and complete transfer of OAM to the BEC, bringing the condensate into a vortex state. Unlike the Raman coupled two-pulse pumping scheme of Refs.~\cite{AndersonPRL2006,WrightPRA2008} this scheme yields several orders of magnitude larger vortex/no-vortex recoil ratio. In addition, the recoiled atoms remain in the same internal state~\cite{WrightPRA2008}.

\begin{figure}[t]
\includegraphics[width=3.35in]{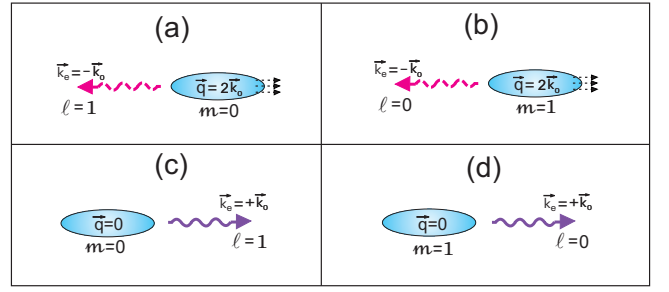}
\caption{(Color online) Cigar-shaped BEC initially in the state $m=0,\vec{q}=0$ and illuminated by
a strong LG-mode laser of  wave vector $\hbar \vec{k}_0$ and in the OAM mode $\ell=1$ and. Superradiant scattering is predominantly along the end-fire modes with wave vector $ \bf{k}_e=\pm {\bf k}_0$. Four possible end-fire modes and the corresponding matter-wave side-modes are shown in (a)-(d).  }
\label{fig1}
\end{figure}

We consider a cigar-shaped BEC in a non-rotating elongated trap, pumped along the long condensate axis ($z$-axis) by a far off-resonant intense laser field of momentum ${\bf k}_0$, see Fig.~\ref{fig1}. For a Fresnel number close to unity the atoms collectively recoil into well-defined momentum states (side-modes), while the scattered light is predominantly along two end-fire modes propagating along $\pm z$~\cite{MeystrePRL1999-MustecapPRA2000}.

After adiabatic elimination of the excited states, the effective Hamiltonian describing the interaction of the condensate atoms of Bohr frequency $\omega_a$ with a far-off resonant optical field detuned from the atomic transition by $\Delta=\omega_a-\omega_0$ is
\begin{eqnarray}
&\hat{\cal H} &=  \int d^3\mathbf{r} \hat{\psi}^{\dagger}({\bf r}) \hat{H}_{\rm g}({\bf r}) \hat{\psi}({\bf r}) + \sum_\ell \int d^3{\bf k} \hbar\omega_k \hat{a}_{{\bf k},\ell}^{\dagger} \hat{a}_{\bf{k},\ell} \nonumber \\
&+&\sum_{\ell \ell'} \int d^3{\bf r} d^3{\bf k} d^3{\bf k}'\tilde{g}_{\ell \ell'}({\bf k},{\bf k}';{\bf r}) \hat{\psi}^{\dagger}({\bf r}) \hat{a}^{\dagger}_{\mathbf{k},\ell} \hat{a}_{{\bf k}',\ell'} \hat{\psi}({\bf r}),
\label{eq:Hamilt2}
\end{eqnarray}
where $\hat{H}_{\rm g}(\mathbf{r}) $ is the atomic Hamiltonian, $\hat{\psi} (\bf{r})$ and the annihilation operator for atoms in their electronic ground state, and $\hat{a}_{\mathbf{k},\ell}$ are optical field mode annihilation operators, the indices $\ell, \ell'=0,\pm 1$ labeling the angular momentum of the optical modes \cite{AllenPRA1992}.

The effective coupling coefficients 
\begin{equation}
\tilde{g}_{\ell \ell'}({\bf k},{\bf k}';{\bf r})=-\frac{\hbar g^*({\bf k}) g({\bf k}')}{\Delta}  \Phi_{{\bf k},\ell}^*({\bf r})
\Phi_{{\bf k'},\ell'}({\bf r})
\end{equation}
are determined by the single atom-photon dipole matrix element $g(\bf{k}) $. Here, $\Phi_{\bf{k},\ell}(\bf{r})$ are the mode functions of the light field with wave number ${\bf k}$ and angular momentum $\hbar \ell.$ For LG modes we have
\begin{equation}
\Phi_{\mathbf{k},\ell}(\mathbf{r}) = \frac{1}{\sqrt{\pi}} \left(\frac{r}{w_\ell}\right)^\ell e^{-r^2/2w_\ell^2}e^{i \ell \phi}e^{ik z} \text{,}
\label{eq:LGmodes}
\end{equation}
of width $w_\ell$ which carries $\ell \hbar$ units of OAM along z-axis.

It is known that superradiance in an atomic BEC can be modeled by an effective Hamiltonian that only includes the dominant end-fire modes of the light field and the associated matter-wave side-modes~\cite{MeystrePRL1999-MustecapPRA2000,MustecapPRA2009,MeystrePRA2007}. We can then proceed by expanding the condensate field operator in terms of side modes \cite{MeystrePRL1999-MustecapPRA2000} as
\begin{equation}
\hat{\psi}_{\rm g}(\mathbf{r})=\sum_{m,q_z} \varphi_{m,q_z}(\mathbf{r}) \hat{c}_{q_z,m},
\label{eq:sidemode-expansion}
\end{equation}
where $\varphi_{m,q_z}(\mathbf{r})=\varphi_{m}(r)e^{i m\phi}e^{iq_z z}$ is the eigenfunction of  an atom with recoil momentum $q_z$ in the $z$-direction and vortex charge $m$. It is given by the solution of the equation $\hat{H}_m(r)\varphi_m(r)=\Omega_m\varphi_m(r)$ where
\begin{equation}
\hat{H}_m({\bf r})=\frac{-\hbar^2}{2M}\left(\frac{d^2}{dr^2}+\frac{1}{r}\frac{d}{dr}\right)+
\frac{m^2\hbar^2}{2Mr^2}+V_t(\bf{r}),
 \end{equation}
and $V_t$ is the trap potential. For a harmonic trap, these states are similar to the LG modes
\cite{MeystrePRA2007}
\begin{equation}
\varphi_m(r)=\frac{1}{\sqrt{m!\pi w}} \left(\frac{r}{w}\right)^m e^{-r^2/2w^2}, %
\label{eq:vortex-eigenstates}
\end{equation}
where $w$ is the radial width of the condensate.

Focusing on first-order scattering processes reduces then the Hamiltonian of the system to the simplified form
\begin{eqnarray}
&\hat{{\cal H}}&=\sum_{m,q_z} \epsilon_{m}(q_z)\hat{c}_{q_z,m}^\dagger \hat{c}_{q_z,m} + \sum_\ell \int d^3{\bf k} \hbar\omega_k \hat{a}_{{\bf k},\ell}^{\dagger} \hat{a}_{{\bf k},\ell} \nonumber \\
&+&\sum_{\ell,m,q_z} \int dk_z  g_{\ell m}(k_z,q_z) \hat{c}_{q_z,m}^\dagger {\hat a}_{k_z,1}^\dagger \hat{a}_{k_0,1}^{(L)} \hat{c}_{0,0} + {\rm h.c.},
\label{eq:Hamilt3}
\end{eqnarray}
where $\epsilon_{m}(q_z)=\hbar(\Omega_m + \omega_R(q_z))$,  $\hat{a}_{k_0,1}^{(L)}$ is the LG pump laser mode,  $\hat{c}_{0,0}$ describes the initial condensate, and
\begin{eqnarray}
&&g_{\ell m}(k_z,q_z)=-\hbar\frac{g^*(k_z)g(k_0)}{\Delta}\delta_{\ell+m,1}\nonumber\\
&\times&\int d^3{\bf r} \varphi_{q_z,m}^*({\bf r}) \Phi_{k_z,\ell}^*({\bf r})
\Phi_{k_0,1}({\bf r})\varphi_{0,0}(\bf{r}). 
\label{eq:coupling-coef}
\end{eqnarray}

When driven by a Gaussian beam of wave vector ${\bf k}_0$, superradiant scattering occurs mainly in two counter-propagating end-fire modes of wave vectors $\mathbf{k}\simeq\pm\mathbf{k}_0$, where ${\bf k}_0$ is the wave vector of the pump photon. \cite{JuntaoLiPLA2008,Deng}, with the initial condensate coupled to two dominant matter-wave side modes of momenta $\mathbf{q}=0$ and $\mathbf{q}=2\mathbf{k}_0$. With a LG  driving field, in contrast, four end-fire optical modes are excited,  with $\mathbf{k}\simeq\pm\mathbf{k}_0$ and  $\ell=0$ or 1, accompanied by four matter-wave side modes $\mathbf{k}\simeq\pm\mathbf{k}_0$ with $m=1$ or 0, the optical end-fire modes with $\ell =0$ being coupled to matter-wave modes with $m = 1$ due to conservation of angular momentum (see Fig.~\ref{fig1}). 

The dynamics of scattering from a superfluid BEC fundamentally differs in the two regimes. If the recoil energy $\left(\hbar\omega_R(q)=\hbar^2 q^2/2M\right)$ is much smaller than the interaction energy (per atom), single particle excitations are highly suppressed \cite{KetterlePRL1999,WallsPRL1996,WrightPRA2008}. The collectivity, induced by the interparticle interaction, does not allow atoms to recoil one by one, i.e. induction of vortex only after a critical rotation frequency \cite{PethickSmith} in rotating trap. In this {\it mean-field} regime, photons are scattered by only the quasiparticle excitations and BEC is described by an order parameter $\psi(\mathbf{r})$. For relatively {\it large recoil energy} \cite{WallsPRL1996}, condensed atoms behave as non-interacting ones and photon scattering is essentially due to
individual atoms one by one \cite{AndersonPRL2006,WrightPRA2008,ketterle,JuntaoLiPLA2008,Deng}.

In the large recoil energy regime, rotatory forward scattering (Fig. 1d)
of LG laser is neglected on the basis that
(i) it cannot transfer sufficient linear momentum and excitation energy \cite{PS1,PS2}
and  (ii) such low-energy scattering are suppressed by the structure factor
of the BEC \cite{KetterlePRL1999,dengPRA2010}.
In the mean-field (smaller recoil frequency) regime,
where single atom scattering is already forbidden \cite{KetterlePRL1999,dengPRA2010,WrightPRA2008},
all processes Fig. 1(a-d) included.

Since the large recoil energy regime corresponds to a series of well-known experiments \cite{JuntaoLiPLA2008,Deng,ketterle}, we investigate it first. Keeping only relevant modes results in the effective Hamiltonian $(\hbar=1)$
\begin{eqnarray}
\hat{{\cal H}}_{\rm int} &=& -\left( g_1 \hat{c}_{2k_0,0}^\dagger \hat{a}_{-k_0,1}^\dagger \hat{a}_{k_0,1}^{\phantom{,}\scriptscriptstyle{(L)}} \hat{c}_{0,0} + \text{h.c.} \right)\nonumber \\ 
&-&\left( g_2 \hat{c}_{2k_0,1}^\dagger
\hat{a}_{-k_0,0}^\dagger \hat{a}_{k_0,1}^{(L)} \hat{c}_{0,0} + \text{h.c.} \right) + \hat{{\cal H}}_{\rm d}\text{,}
\label{eq:hamiltonian}
\end{eqnarray}
where $\hat{c}_{2k_0,0}$, $\hat{c}_{2k_0,1}$, $\hat{a}_{-k_0,1}$ and $\hat{a}_{-k_0,0}$ are the matter-wave side modes and optical end-fire modes illustrated in Figs.~\ref{fig1}a and \ref{fig1}b. Here $g_1\equiv g_{10}=w_1^2w^2/2\pi^2\left(w^2+w_1^2\right)^2$
and $g_2\equiv g_{01}=(w_1/w) g_1$ can be calculated from Eq.~(\ref{eq:coupling-coef}), and $\hat{{ \cal H}}_{\rm d}=- 2g_0 \hat{c}_{0,0}^\dagger \hat{a}_{k_0,1}^{\dagger, (L)} \hat{a}_{k_0,1}^{\phantom{,}\scriptscriptstyle{(L)}}\hat{c}_{0,0}$ is a diagonal light shift term with $g_0=g_1$. The free-field terms have been eliminated through a second rotating frame transformation after moving to a co-rotating frame at the laser frequency~\cite{MustecapPRA2009}. The first term in Eq.~(\ref{eq:hamiltonian}) yields normal (non-rotatory) SR, while the second term induces vortex excitations in the BEC and results in rotatory SR.

For times short enough that the depletion of the initial condensate can be ignored we can make the substitution ${\hat c}_{0,0} \rightarrow \sqrt{N}$, where $N$ is the number of condensed atoms. In that limit the side-mode population dynamics can be treated analytically~\cite{MeystrePRL1999-MustecapPRA2000}, giving $d\hat{c}_{2k_0,m}/dt\simeq G_m N \hat{c}_{2k_0,m}/2$, with $G_m\propto g_{m+1}^2$. For  $g_2/g_1=w_1/w \gtrsim 1$, the initial exponential growth of vortex side-mode occupation from initial fluctuations is faster than that of the non-vortex side mode. This suppression of normal SR relative to rotatory SR, in the large recoil energy regime, can be thought of as a topological vortex phase transition. For example, a typical ratio $g_2/g_1=2$ \cite{AndersonPRL2006} would result in vortex/no-vortex side mode population ratio of   $3\times 10^4$ for $N=10^6$. We note that the condition $w_1 \gtrsim w$, is consistent with OAM transfer into classical objects \cite{ONeilPRL2002} and with the two-pump Raman vortex excitation protocol in BEC
\cite{AndersonPRL2006}. In these cases, however, the resulting vortex/no-vortex population ratio increases \textit{linearly} with  $(w_1/w)^2$, in contrast to the \textit{exponential} growth characteristic of the present situation.

To investigate the \textit{small recoil energy} regime, we include the effects of two-body collisions and we invoke the mean-field approximation whereby the field operators are replaced by  $c$-numbers,  specifically $\hat{\psi}\rightarrow \psi$, $\hat{a}_{-k_0,1}\rightarrow\alpha_1$, $\hat{a}_{-k_0,0}\rightarrow\alpha_2$, $\hat{a}_{k_0,0}\rightarrow\alpha_3$, and $\hat{a}_{k_0,1}^{(L)}\rightarrow \alpha_{L}$, without side-mode expansion (\ref{eq:sidemode-expansion}). With these approximations, the Heisenberg equations of motion reduce to
\begin{eqnarray}
 i \dot{\psi}&=&\hat{h}_{\rm g}\psi -2U_0 \Big[ \left|\alpha_{L}\right|^2 \left|\Phi_{\rm k_0,1}\right|^2 +\left|\alpha_1\right|^2 \left|\Phi_{\rm -k_0,1}\right|^2  \nonumber \\
&+& \left|\alpha_2\right|^2 \left|\Phi_{\rm -k_0,0}\right|^2 +\left|\alpha_3\right|^2 \left|\Phi_{\rm k_0,0}\right|^2 \Big]  \psi  \nonumber \\
& -& U_0\alpha_{L} \left [ \alpha_1^*\Phi_{\rm -k_0,1}^*\Phi_{\rm k_0,1} +\alpha_2^*\Phi_{\rm -k_0,0}^*\Phi_{\rm k_0,1}  \right .\nonumber \\
&+& \left .\alpha_3^*\Phi_{\rm k_0,0}^*\Phi_{\rm k_0,1} + {\rm c.c.} \right ]  \psi +g_S \left|\psi\right|^2  \psi,\\
i \dot{\alpha}_1 &=& -\Delta_1 \alpha_1
-2U_0 I_{-,-}^{1,1} \alpha_1 - U_0\alpha_{L} I_{-,+}^{1,1}, \label{eq:alpha1} \\
i \dot{\alpha}_2 &=& -\Delta_2 \alpha_2 -2U_0 I_{-,-}^{0,0} \alpha_2 - U_0\alpha_{L} I_{-,+}^{0,1}, \label{eq:alpha2} \\
i \dot{\alpha}_3 &=& -\Delta_3 \alpha_3 -2U_0 I_{+,+}^{0,0} \alpha_3 - U_0\alpha_{L} I_{+,+}^{0,1}, \label{eq:alpha3}.
\end{eqnarray}
Here $U_0=g_1^2/\Delta$,  $g_S=4\pi\hbar a_s/m$,  $a_s$ being the $s$-wave scattering length, $-\Delta_{1,2,3}$ are the end-fire mode frequencies in the rotating frame at frequency $\omega_0$, and
\begin{equation}
I_{\alpha,\beta}^{m,m'} = \int d^3{\bf r} \Phi_{\alpha k_0,m}^*({\bf r}) \Phi_{\beta k_0,m'}({\bf r}) \left|\psi({\bf r},t)\right|^2,
\end{equation}
where $\alpha$ and $\beta=\pm 1$ label the sign of the wave vectors of amplitude  $k_0$.

\begin{figure}
\includegraphics[width=3.3in]{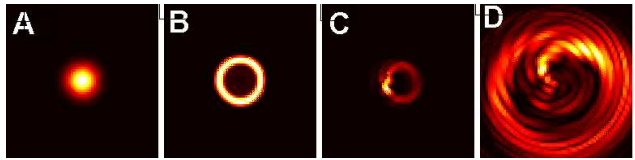}
\caption{(Color online) Snapshots illustrating the onset of BEC SR in a BEC driven by a LG laser beam. (a) The BEC starts in a Gaussian state; (b) Onset of normal SR, with the  condensate density mapping the LG intensity profile. (c) Just before the onset of rotatory SR the axial $\phi$-symmetry of the condensate is broken, with a higher density near the $\phi \simeq \pi$ angle. (d) Rotatory SR induces a vortex topological phase transition in the BEC. }
\label{fig2}
\end{figure}

\begin{figure}
\includegraphics[width=3.65in]{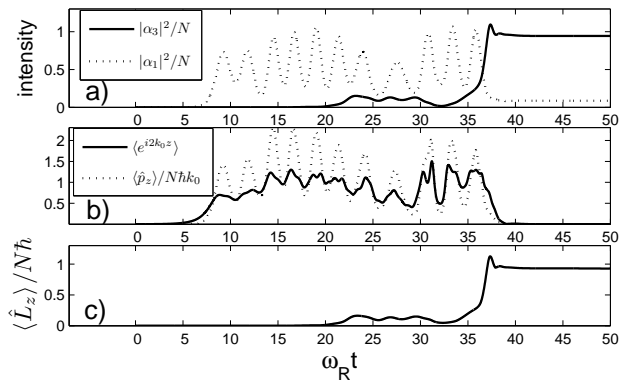}
\caption{ Time evolution of  the condensate for the parameters of Fig.~\ref{fig2}.
(a) Mean photon number in the Gaussian end-fire mode $|\alpha_3|^2$ and in the LG end-fire mode. (b)  Linear momentum transfer ($\langle p_z\rangle\propto |\alpha_1|^2$)
and expectation value of the two-photon momentum recoil operator $\langle e^{2ik_0z}\rangle$;
(c), Expectation value of the normalized angular momentum $\langle {\hat L}_z\rangle/N \hbar$. }
\label{fig3}
\end{figure}

Figures \ref{fig2} and \ref{fig3} summarize the results of a typical numerical solution of these equations. In this example $w_1=1.2w$, $NU_0=10$, $\eta=\alpha_L\sqrt{N}U_0=15$, $\Delta_1=1$, $\Delta_{2}=1.001$, and $\Delta_3=10^{-3}$ in units of the recoil frequency $\omega_R$. These values are comparable to those of Ref.~\cite{ketterle,NagyEPJD2008-DomokosPRL2002}. (We note that the energy scale of light shift is several orders of magnitude larger than all other terms, however its effect is transient in and of no significant importance for the time scales under consideration. )

The results of our simulations are perhaps most dramatically illustrated in Fig.~\ref{fig2}, which shows four snapshots of the transition from an initial condensate in the ground state of the harmonic trap to a condensate with a density profile that maps the intensity profile of the LG driving field, but still with zero average angular momentum, and subsequently to a rotatory condensate carrying one quantum of angular momentum per atom, that is, a condensate in a vortex state. 

The dynamics of the transition are shown in more detail in Fig.~\ref{fig3}, which plots the evolution of key observables during the SR transition to the vortex phase: For the parameters of these simulations the normal SR PT happens at $t\simeq 8/\omega_R$ (Fig. \ref{fig2}b). It coincides with a peak in the intensity $|\alpha_1|^2$ (Fig. \ref{fig3}a of the LG end-fire mode $\hat{a}_{-k_0,1}$. Spatial order sets in along the $z$-direction, see  (Fig. \ref{fig3}b), as evidenced by the non-zero value of $\langle e^{2ik_0z}\rangle$ and the fact that the condensate acquires linear momentum $\langle\hat{p}_z\rangle/\hbar k_0=2|\alpha_1|^2$ as a result of momentum conservation. In this early phase, there is no OAM transfer to the condensate happens as both the laser and the end-fire modes have equal winding number $\ell=1$.

This early dynamics is followed by the onset of rotatory SR.  As a precursor to that transition the condensate first exhibits an axial symmetry breaking both in its phase and density, as illustrated in Fig.~\ref{fig2}c. This is the first indication of OAM transfer from the optical field to the atoms~\cite{PS3}. Rotatory SR reveals itself with a sudden increase in the intensity $|\alpha_3|^2$ of the Gaussian end-fire mode $\hat{a}_{-k_0,0}$, shown in Fig.~\ref{fig3}a, and a decrease in the Laguerre-Gaussian mode $|\alpha_1|^2.$ At this point the angular momentum is transferred fully from the optical field to the matter wave, establishing a vortex in the BEC (Fig.~\ref{fig2}d), the  BEC gaining $\langle\hat{L}_z\rangle=N\hbar$ (Fig.~\ref{fig3}c) of orbital angular momentum. Comparison of Figs.~\ref{fig3}a and \ref{fig3}c indicates that the total OAM of the system is conserved in this transition, during which the process the linear momentum along $z$ is transferred back to the optical field, as can be seen from Figs.~\ref{fig3}a and \ref{fig3}b.

We finally remark that we found numerically that the critical pump rates ($\eta$) are different for the onset of non-rotatory ($\eta_1=9.5$) \cite{NagyEPJD2008-DomokosPRL2002,EsslingerNature2010,EmaryPRE2003}, and rotatory SR ($\eta_2=12.4$): Below $\eta_1$ no SR scattering occurs. Between the two values ($\eta_1<\eta<\eta_2$) only normal SR takes place, and rotatory SR occurs for above $\eta_2$.

Summarizing, we have studied theoretically the mutual induction of a topological vortex phase transition in a BEC and a rotatory superradiant phase transition-like . The cooperative nature of SR from an incident LG laser beam allows for a sudden transfer of a large amount of OAM into a condensate, bringing it into a vortex state. For most current condensate SR experiments, non-rotatory SR can be exponentially suppressed by rotatory SR if the transverse width of the LG pump laser is larger than the condensate transverse width.

We gratefully thank Pierre Meystre for many illuminating discussions and help with the manuscript. \"O.E.M. and M.E.T. acknowledge support by D.P.T. (T.R. Prime Ministry State Planning Organization) National Quantum Cryptology Center. M.E.T. acknowledges support from TUBITAK 2219 program. \"O.E.M. acknowledges support from TUBITAK Project No. 109T267.


%
%
%

\begin{thebibliography}{99}
\bibitem{Beth} R.A. Beth, Phys. Rev. \textbf{48}, 471 (1935); {\it ibid} \textbf{50}, 115 (1936).

\bibitem{cohen} C. Cohen-Tannoudji, J. Dupont-Roc, and G. Grynberg,
Photons and Atoms (Wiley, New York, 1989).

\bibitem{AllenPRA1992} L. Allen \textit{et al.}, Phys. Rev. A \textbf{45}, 8185 (1992).

\bibitem{ZeilingerNature2001} A. Mair, A. Vaziri, G. Weihs, and A.
Zeilinger, Nature \textbf{412}, 313 (2001).

\bibitem{MolinaPRL2002} G. Molina-Terriza, J.P. Torres, and L.
Torner, Phys. Rev. Lett. \textbf{88}, 013601 (2001).

\bibitem{MolinaPRL2004-MolinaNaturePhys2007} G. Molina-Terriza \textit{et al.},
Phys. Rev. Lett. \textbf{92}, 167903 (2004).; G. Molina-Terriza, J.P. Torres, and
L. Torner, Nature Phys. \textbf{3}, 305 (2007).

\bibitem{KazslikowskiPRL2000} D. Kaszlikowski \textit{et al.},
Phys. Rev. Lett. \textbf{85}, 4418 (2000).

\bibitem{CollinsPRL2002} D. Collins, N. Gisin, N. Linden, S.
Massar, and S. Popescu, Phys. Rev. Lett. \textbf{88}, 040404
(2002).

\bibitem{TabosaPRL1999} J.W.R. Tabosa and D.V. Petrov, Phys. Rev.
Lett. \textbf{83}, 4967 (1999).

\bibitem{AndersonPRL2006} M.F. Andersen \textit{et al.},
Phys. Rev. Lett. \textbf{97} 170406 (2006).

\bibitem{WrightPRA2008} K.C. Wright, L.S. Leslie, and N.P.
Bigelow, Phys. Rev. A \textbf{77} 041601(R) (2008).

\bibitem{MorettiPRA2009} D. Moretti, D. Felinto, and J.W.R. Tabosa, Phys. Rev.
A\textbf{79}, 023825 (2009).


\bibitem{PethickSmith} C.J. Pethick and H. Smith, Bose-Einstein
Condensation in Dilute Gases (Cambrige University Press, UK, 2002)
p. 250.

\bibitem{MarzlinPRL1997} K.-P. Marzlin, W. Zhang, and E.M. Wright, Phys. Rev.
Lett. \textbf{79} 4728 (1997).

\bibitem{MeystrePRA2007} R. Kanamoto \textit{et al.},
Phys. Rev. A \textbf{75}, 063623 (2007).

\bibitem{DuttonPRL2004-ThanvanthriPRA2008} Z. Dutton and J. Ruostekoski, Phys. Rev.
Lett. \textbf{93}, 193602 (2004); S. Thanvanthri, K.T. Kapale, and J.P.
Dowling, Phys. Rev. A \textbf{77}, 053825 (2008).

\bibitem{dicke} R. H. Dicke, Phys. Rev. {\bf 93}, 99 (1954).

\bibitem{skribanowitz} N. Skribanowitz \textit{et al.}, Phys. Rev. Lett. {\bf 30}, 309
(1973).

\bibitem{WangPRA1973-heppAnnPhys} Y.K. Wang and F.T. Hioe, Phys. Rev. A
\textbf{7}, 831 (1973); K. Hepp and E.H. Lieb, Ann. Phys.
\textbf{76}, 360 (1973).

\bibitem{EmaryPRE2003} C. Emary and T. Brandes, Phys. Rev. E \textbf{67}, 066203
(2003).

\bibitem{ketterle} S. Inouye \textit{et al.}, Science \textbf{285},
571 (1999).

\bibitem{JuntaoLiPLA2008} J. Li \textit{et al.}, Phys. Lett. A \textbf{372} 4750
(2008).

\bibitem{Deng} L. Deng \textit{et al.}, Phys. Rev. Lett. \textbf{104}, 050402 (2010).

\bibitem{bonifacioexp} L. Fallani {\it et al.} Phys. Rev. A {\bf 71}, 033612 (2005).

\bibitem{ketterleRaman} D. Schneble \textit{et al.}, Phys. Rev. A \textbf{69}, 041601(R) (2004).

\bibitem{MeystrePRL1999-MustecapPRA2000} M. G. Moore and P. Meystre, Phys. Rev. Lett.
\textbf{83}, 5202 (1999); O.E. Mustecaplioglu and L. You, Phys. Rev.
A \textbf{62}, 063615 (2000).

\bibitem{zobaySpatial} O. Zobay and G. M. Nikolopoulos, Phys. Rev. A {\bf 73}, 013620 (2006).

\bibitem{meystreRamanSR} H. Uys and P. Meystre, Phys. Rev. A {\bf 75}, 033805 (2007).

\bibitem{EsslingerNature2010} K. Baumann, C. Guerlin, F. Brennecke,
and T. Esslinger, Nature \textbf{464}, 1301 (2010).

\bibitem{NagyEPJD2008-DomokosPRL2002} D. Nagy, G. Szirmaia, and P. Domokos, Eur. Phys. J. D \textbf{48}, 127
(2008); P. Domokos and H. Ritsch, Phys. Rev.
Lett. \textbf{89}, 253003 (2002).

\bibitem{KetterlePRL1999} D.M. Stamper-Kurn \textit{et al}., Phys.
Rev. Lett. \textbf{83}, 2876 (1999).

\bibitem{WallsPRL1996} R. Graham and D. Walls, Phys. Rev. Lett.
\textbf{76}, 1774 (1996).

\bibitem{MustecapPRA2009} M.E. Ta\c{s}g\i n
\textit{et al}., Phys. Rev. A \textbf{79}, 053603 (2009).

\bibitem{PS1} In Fig. 1d, excitation energy of the side-mode is
many orders of magnitude smaller than the atomic interaction energy.
Recoil into this side-mode can take place \cite{KetterlePRL1999,WallsPRL1996} only if
all ($N$) of the condensate atoms are involved, similar to vortex formation in a rotating trap.
Such a process transfers $\sim N\hbar$ OAM {\it at once}, thus is
prohibited by the vanishingly small probability
compared to other high excitation energy side-modes Figs. 1a and 1b.

\bibitem{PS2} Recoil into side-mode Fig. 1c can happen freely,
since both initial and final modes are the same ($\mathbf{q}=0$,$m=0$).

\bibitem{dengPRA2010} L. Deng and E.W. Hagley, Phys. Rev. A {\bf 82}, 053613 (2010).

\bibitem{ONeilPRL2002} A.T. O'Neil \textit{et al.}, Phys. Rev. Lett. \textbf{88}, 053601 (2002).

\bibitem{PS3} Analogously, rotation experiments using classical objects 
reveal \cite{ONeilPRL2002} higher OAM is transferred to materials with higher $\phi$-anisotropic density.

\end{thebibliography}
\end{document}